\documentclass[prb,twocolumn,english,footinbib]{revtex4-2}
\usepackage{amsmath,amssymb,mathrsfs}
\usepackage{amsfonts,bm}
\usepackage{amsmath}
\usepackage{amssymb}
\usepackage{amsthm}
\usepackage{graphicx}
\usepackage{graphics}
\usepackage{subfigure}
\usepackage{color}
\usepackage{bm}
\usepackage{hyperref}
\usepackage{rotating}
\usepackage{comment}
\usepackage[colorinlistoftodos]{todonotes}
\usepackage{esint}
\usepackage{xfrac}
\usepackage{natbib}
\usepackage{upgreek}

\newcommand{\be}{\begin{equation} }
\newcommand{\ee}{\end{equation} }
\newcommand{\ba}{\begin{eqnarray} }
\newcommand{\ea}{\end{eqnarray} }

\newcommand{\bpm}{\begin{pmatrix}}
\newcommand{\epm}{\end{pmatrix}}
\newcommand{\bmm}{\begin{matrix}}
\newcommand{\emm}{\end{matrix}}

\newcommand{\la}{\label}
\newcommand{\p}{\partial}

\newcommand{\bea}{\begin{eqnarray}}
\newcommand{\eea}{\end{eqnarray}}

\makeatother

\usepackage{babel}

\begin{document}


\title{Hele-Shaw flow for parity odd three-dimensional fluids}

\author{Dylan Reynolds}
\author{Gustavo M. Monteiro}
\author{Sriram Ganeshan}

\affiliation{Department of Physics, City College, City University of New York, New York, NY 10031, USA }
\affiliation{CUNY Graduate Center, New York, NY 10031}

\date{\today}


\begin{abstract}

The Hele-Shaw cell is a device used to study fluid flow between two parallel plates separated by a small gap. The governing equation of flow within a Hele-Shaw cell is Darcy's law, which also describes flow through a porous medium. In this work, we derive a generalization to Darcy's law starting from a three dimensional fluid with a parity-broken viscosity tensor with no isotropy. We discuss the observable effects of parity odd fluids in various physical setups relevant to Hele-Shaw experiments, such as channel flow, flow past an obstacle, bubble dynamics, and the Saffman-Taylor instability. In particular, we show that when such a fluid is pushed through a channel, a transverse force is exerted on the walls, and when a bubble of air expands into a region of such fluid, a circulation develops in the far field, with both effects proportional to the parity odd viscosity coefficients. The Saffman-Taylor stability condition is also modified, with these terms tending to stabilize the two fluid interface. Such experiments can in principle facilitate the measurement of parity odd coefficients in both synthetic and natural active matter systems.

\end{abstract}


\maketitle


\section{Introduction} \la{sec:intro}

In three dimensions, isotropic fluids possess three independent viscosity coefficients, namely shear viscosity, bulk viscosity, and rotational viscosity. Shear viscosity introduces friction between adjacent fluid layers that flow with a relative velocity differential, bulk viscosity provides resistance to compression or expansion of the fluid, and rotational viscosity gives rise to torque when the fluid vorticity is non-zero. 
For incompressible flows, the bulk viscosity term vanishes and the fluid pressure is entirely determined by the flow, i.e., it does not come from an equation of state. We will restrict ourselves to incompressible flows in this work. 

Both shear and rotational viscosity break time reversal symmetry due to their dissipative nature while preserving parity symmetry~\cite{landau1987fluid}. Viscosity coefficients that break parity in three dimensions can only be realized in anisotropic systems. This is in contrast to 2D systems where there exists parity breaking viscosity coefficients that are consistent with isotropy. Odd viscosity is an example of such coefficient, and it has been investigated extensively in both classical and quantum two-dimensional systems~\cite{avron1995viscosity, avron1998odd, tokatly2006magnetoelasticity,tokatly2007new,tokatly2009erratum, read2009non,haldane2011geometrical,haldane2011self,hoyos2012hall, bradlyn2012kubo, yang2012band,abanov2013effective,hughes2013torsional, hoyos2014hall, laskin2015collective, can2014fractional,can2015geometry,klevtsov2015geometric,klevtsov2015quantum, gromov2014density, gromov2015framing, gromov2016boundary,alekseev2016negative, scaffidi2017hydrodynamic, pellegrino2017nonlocal, berdyugin2019measuring}. Parity breaking flows in three dimensions have been considered in 3D plasmas in the presence of a magnetic field~\cite{landau1987fluid,Monteiro2018nonresistivite}, and systems with polyatomic molecules~\cite{korving1966transverse, knaap1967heat, korving1967influence} . Recent work of Khain et al~\cite{vitelli2021long} study the effects of parity-violating and non-dissipative viscosities for three dimensional Stokes flows. For active matter systems the parity violating coefficients stem from the relaxation of the fluid's intrinsic angular momentum dynamics ~\cite{banerjee2017chiral, lubensky2021anal, monteiro2021hamiltonian} . 

In general, for incompressible fluids with no symmetry whatsoever, the viscosity tensor is a daunting object with 64 independent coefficients. In this paper, we show that despite this complexity, a remarkable simplification happens when such a fluid is placed in a Hele-Shaw (HS) cell, a physical setup where the fluid is confined in a small separation between two plates. The governing equations of an isotropic fluid in a HS geometry is given by Darcy's law,
\begin{align}
    V_i=-\frac{h^2}{12 \eta} \partial_i P,
    \label{eq:darcy}
\end{align}
where $V_i(x,y)$ is the gap averaged 2D flow between the two plates separated by a small separation $h$, $\eta$ is the fluid shear viscosity, and $P(x,y)$ is the fluid pressure. The form of Darcy's law given in Eq.~(\ref{eq:darcy}) universally applies to fluids flowing through a porous medium~\cite{batchelor2000flow, smith1969flow}. It is also analogous to Ohm's law in isotropic media, where pressure is replaced by the scalar electric potential and the constant $-\frac{h^2}{12 \eta}$ is replaced by the conductivity divided by the charge density~\footnote{In fact, Darcy's law also manifests as Fourier's law of heat conduction and Fick's law of diffusion.}. For a general anisotropic incompressible fluid, we show that flow in a Hele-Shaw cell is governed by a modified Darcy's law that takes the simple form,
\begin{align}
V_i=-\frac{h^2}{12}(\mathfrak{y}^{-1})_{ij}\partial_j P, \la{eq:gen-Darcy}
\end{align}
where we have defined the matrix $\mathfrak{y}$ in terms of the components of the full three-dimensional rank 4 viscosity tensor $\eta_{ijkl}$,
\begin{align}
    \mathfrak{y}=\bpm \eta_{xzzx}\,\,\, \eta_{xzzy}\\ \eta_{yzzx}\,\,\, \eta_{yzzy} \epm . \la{eta-matrix}
\end{align}
Without specifying the symmetries of the underlying viscosity tensor, the precise form of the coefficients in (\ref{eta-matrix}) cannot be determined. However, the explicit appearance of $z$ indices means that these terms have no 2D counterparts; they are unique to 3D flows. Thus, the odd viscosity coefficient appearing in purely 2D systems does not contribute.

The form of Eq.~(\ref{eq:gen-Darcy}) is analogous to two-phase flows through anisotropic porous media~\cite{dmitriev1998determining}, and Ohm's law with an anisotropic conductivity tensor in two dimensions. However, the coefficients in the fluid case are all transport coefficients associated with a first order gradient expansion. For a fluid with cylindrical symmetry aligned perpendicular to the HS cell, the system further simplifies, since $\eta_{xzzx}=\eta_{yzzy}$ and $\eta_{xzzy}=-\eta_{yzzx}$. Assuming this symmetry, we consider several examples of typical HS setups, such as single fluid channel flow, an expanding bubble, and the Saffman-Taylor instability, and derive observable consequences of the parity breaking terms for HS flows. 

The broad applicability of HS flows means that our analysis can easily be adapted to many relevant physical situations. For example, HS cells have been used to study the behavior of active matter and micro swimmers~\cite{tsang2015circ, miles2019active}, and the analysis done here could reveal the parity odd nature of the fluid. This would enable measurement of these coefficients for many complex anisotropic fluids. Future experimental work could also focus on confinement of microrollers and colloidal magnetized particles to a HS cell in order to probe their parity odd behavior~\cite{driscoll2017rollers, soni2018free}.

This paper is organized as follows. In Section \ref{sec:general} we derive Darcy's law in the presence of a general viscosity tensor. In Section \ref{sec:cylinder} we impose cylindrical symmetry, which is used in the rest of the main paper, while in Appendix \ref{sec:aniso} we show these results can be extended to the case of a general viscosity tensor by a simple coordinate transformation. In Section \ref{sec:observe} we discuss the observable modifications to results involving flow in a channel, force on an obstacle, expanding bubble, and free surface stability. We end the paper with a discussion on possible microscopic magnetic systems akin to ferrofluids that can serve as a platform to realize some of the physics discussed in this paper.

\section{Parity Odd Three-dimensional fluids in a Hele-Shaw setup} \la{sec:general}

The starting point of our hydrodynamic system is the equations governing local conservation of momentum and mass. For an incompressible fluid they can be written in terms of the flow velocity $v_{i}$ and constant density $\rho$ as
\begin{align}
\rho \left(\partial_{t}v_{i}+v_{j}\partial_{j}v_{i}\right)=\partial_{j}T_{ij}+f_{i},\quad \partial_{i}v_{i}=0 \la{eq:ns}.
\end{align}
The external force density $f_{i}$ is assumed to come from a uniform gravitational field in the negative $y$ direction, however the following analysis can be extended to an arbitrary external force. For a completely general viscosity tensor, which we will assume to be uniform throughout the fluid, the stress tensor $T_{ij}$ takes the form
\begin{align}
T_{ij}= -P\delta_{ij} + \eta_{ijkl}\partial_{k}v_{l},
\end{align}
where $P$ is the pressure. In this paper we will ignore any thermal effects, so energy conservation comes automatically.

\begin{figure}
\centering
\includegraphics[scale=0.27]{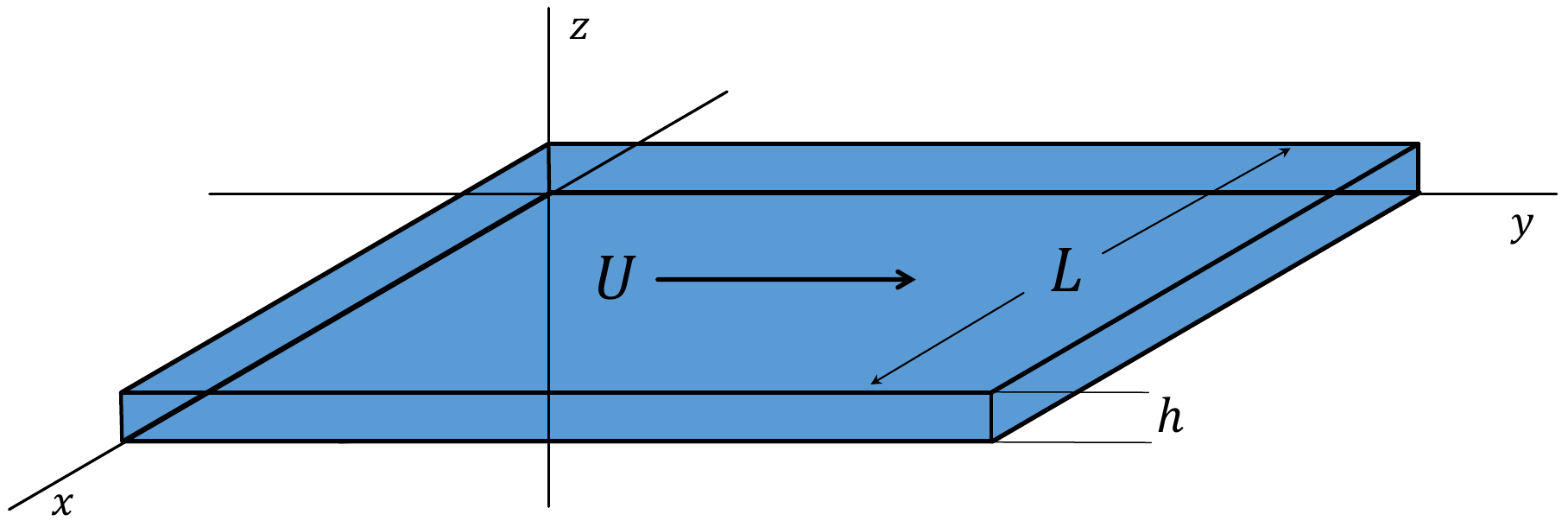}
\caption{Geometry of the Hele-Shaw cell. Fluid is confined between horizontal plates at $z=0$ and $z=h$. Length and velocity scales in the $xy$ plane are of size $h/\epsilon$ and $\nu/h$, respectively.}
\label{fig:schematic1}
\end{figure}

The fluid is now confined between two parallel plates aligned with the $xy$ plane, having a separation $h$ (see Fig~\ref{fig:schematic1}). This introduces a characteristic length scale to the system, and we can derive Darcy's law by assuming the hydrodynamic variables vary in the $xy$ plane at much larger length scales than the distance $h$ between the two plates. This can be formally introduced by defining 
\be
x=\frac{h}{\epsilon}\bar x\,,\qquad y=\frac{h}{\epsilon}\bar y\,,\qquad z=h\bar z\,,
\ee
where all barred quantities are dimensionless, and $\epsilon\ll 1$. The viscosity tensor introduces another dimensionful parameter to the system. In fact, $\eta_{ijkl}/\rho$ has dimension of $(length)^2/time$, which introduces a characteristic time and velocity scale to the system. Let $\rho \nu$ be a representative component of the viscosity tensor (usually the shear viscosity). The characteristic time and the velocity scale are then given by $h^2/\nu$ and $\nu/h$, respectively. For example, the kinematic shear viscosity of glycerine is approximately $650\,mm^2/s$ at $20^\circ \,C$ \cite{segur1951viscosity} , and for a HS cell with $h=1 \,mm$ this leads to a characteristic time of $0.0015\, s$ and a velocity scale of $0.65\,m/s$. We can then introduce the rest of the scaling by
\begin{align}
    t=\frac{h^2}{\epsilon \nu}\bar t,\quad v_x=\frac{\nu}{h}\bar v_x,\quad v_y=\frac{\nu}{h}\bar v_y,\quad v_z=\frac{\epsilon\nu}{h}\bar v_z.
\end{align}

The gravitational force combines with the pressure term in such a way that $P$ and $\rho g y$ scale the same. Assuming that $\rho g $ scales as $\epsilon^0$, we have
\be
P+\rho g y=\frac{\rho\nu^2}{\epsilon h^2}\bar P.
\ee
With these scalings, the components of Eq.~(\ref{eq:ns}) become
\begin{align}
 -\partial_{\bar{x}}\bar{P} + \frac{\eta_{xzzx}}{\rho \nu}\partial_{\bar{z}}^2\bar{v_x}+ \frac{\eta_{xzzy}}{\rho \nu}\partial_{\bar{z}}^2\bar{v_y}=\mathcal{O}(\epsilon)&\,,\la{eq:scale1}
 \\
-\partial_{\bar{y}}\bar{P} + \frac{\eta_{yzzx}}{\rho \nu}\partial_{\bar{z}}^2\bar{v_x}+ \frac{\eta_{yzzy}}{\rho \nu}\partial_{\bar{z}}^2\bar{v_y}=\mathcal{O}(\epsilon)&\,,\la{eq:scale2}
\\
-\partial_{\bar{z}}\bar{P}=\mathcal{O}(\epsilon)&\,,\la{eq:scale3}
\\
\p_{\bar x} \bar v_x+\p_{\bar y} \bar v_y+\p_{\bar z} \bar v_z=0&\,. \la{eq:incompress}
\end{align}
The solutions for these equations that satisfy the no-slip boundary conditions on the plates are
\begin{align}
\bar{v}_x(\bar{x},\bar{y},\bar{z}) &= 6\bar{z}(1-\bar{z})\bar V_x(\bar{x},\bar{y})\,,
\\
\bar{v}_y(\bar{x},\bar{y},\bar{z}) &= 6\bar{z}(1-\bar{z})\bar V_y(\bar{x},\bar{y})\,,
\\
\bar{v}_z(\bar{x},\bar{y},\bar{z}) &=0\,,
\end{align}
where $\bar V_x$ and $\bar V_y$ are, respectively, the average values of $\bar{v}_x$ and $\bar v_y$ along the $z$ direction. Restoring dimensions, we have
\begin{align}
\frac{h^2}{12} \partial_x (P+\rho g y) + \eta_{xzzx}V_x + \eta_{xzzy}V_y&=0 \,, \la{eq:Darcy-x}
\\
\frac{h^2}{12} \partial_y (P+\rho g y) + \eta_{yzzx}V_x + \eta_{yzzy}V_y&=0 \,, \la{eq:Darcy-y}
\\
\p_x V_x+\p_y V_y&=0\,, \la{eq:incompress2}
\end{align}
where $V_x=\bar V_x\nu/h$ and $V_y=\bar V_y\nu/h$ are the dimension-full average velocities. The above equations can be combined into a single matrix equation,
\be
\p_a (P+\rho g y)=-\frac{12}{h^2} \mathfrak{y}_{ab} V^b, \la{Darcy}
\ee
where $\mathfrak{y}_{ab}$ is given by Eq.~(\ref{eta-matrix}). Splitting $12 \mathfrak y/h^2$ into its symmetric and anti-symmetric pieces, we end up with
\be
\p_a (P+\rho g y)=-\left(\alpha \gamma_{ab}+\beta \epsilon_{ab}\right) V^b, \la{Darcy-anis}
\ee
where $\gamma_{ab}=\gamma_{ba}$, and $\alpha$ is defined such that $\det \gamma=1$.

The matrix $\gamma_{ab}$ contains only parity-even contributions, while the constant $\beta$ contains only parity-odd contributions. The form of both $\gamma_{ab}$ and $\beta$ depend entirely on the particular fluid under consideration, however they contain only the coefficients with some three-dimensional nature. For example, the coefficients $\mu_2$ and $\eta_2^o$ in Khain et al~\cite{vitelli2021long} would contribute to $\beta$, while $\mu_1$ and $\eta_1^o$ would not. Futurmore, both parity-odd coefficients in Robredo et al~\cite{robredo2021tetra} would contribute, as they are 3D in nature. In this work we focus only on the general observable consequences of a non-zero $\beta$, and do not focus on which parity odd coefficients constitute this $\beta$.

\section{Cylindrically symmetric case} \la{sec:cylinder}

In this paper we restrict our focus to systems where the viscosity tensor has cylindrical symmetry along the $z$ axis, and discuss various observable consequences of the parity breaking terms. In Appendix \ref{sec:aniso} we show that our results can be generalized to the anisotropic case, with most of the results unchanged. For a viscosity tensor with cylindrical symmetry along the $z$ axis, Eq.~(\ref{Darcy-anis}) simplifies. The matrix $\gamma_{ab}$ reduces to $\delta_{ab}$, and 
\begin{align}
\alpha &= \frac{12}{h^2}\eta_{xzzx}=\frac{12}{h^2}\eta_{yzzy},\\
\beta &= \frac{12}{h^2}\eta_{xzzy}=-\frac{12}{h^2}\eta_{yzzx}.
\end{align}
In this case, the system is analogous to a 2D electronic flow subjected to a magnetic field $\beta$ pointing in the positive $z$ direction, electrostatic potential given by $-(P+\rho g y)$ and collision relaxation time given by $m_*/(e\alpha)$, where $e$ is the elementary charge and $m_*$ is the effective mass of the electron \footnote{Note that Darcy's law is analogous to Ohm's law for electrical networks, Fourier's law of heat conduction and Fick's law of diffusion. All of these situations mimic flow through a porous medium.}. In this scenario, Eq.~(\ref{Darcy-anis}), along with Eq.~(\ref{eq:incompress2}), imply that the pressure satisfies Laplace's equation, and the average flow is irrotational, that is,
\be
\Delta P=0\,,\qquad \p_xV_y-\p_yV_x=0\,.
\ee
Therefore, the function $V=V_x-iV_y$ is analytic, i.e., it satisfies the Cauchy-Riemann equations. Moreover, since $P(x,y)$ is a harmonic function, we can always define a function $Q(x,y)$ such that $W=P+iQ$ is analytic. In terms of the complex variables $\zeta\equiv x+iy$, $V$, and $W$, Eq.~(\ref{Darcy-anis}) becomes
\be
\frac{d}{d\zeta}(W-i\rho g\zeta)=-\mu V(\zeta), \la{eq:Darcy-complex}
\ee
where we have introduced a complex valued viscosity coefficient $\mu=\alpha+i \beta$. An immediate consequence of $\mu$ having a complex component is that the fluid flows at an angle
\begin{align}
\theta = \arctan\left(\frac{\beta}{\alpha}\right),\la{eq:angle}
\end{align}
relative to the pressure gradient. This is a manifestation of the typical behavior seen in parity-odd two-dimensional systems. For example, in the classical Hall effect, the electric current makes an angle with respect to the electric field.

\section{Observable effects of parity breaking in a Hele-Shaw setup} \la{sec:observe}

In the following, we discuss the effect of parity breaking coefficients relevant to HS experimental setups. We consider the case of single fluid flow in a channel, drag force on an obstacle, bubble dynamics, and the Saffman-Taylor instability problem.

\subsection{Single fluid flow in a channel}

\begin{figure}
\centering
\includegraphics[scale=0.85]{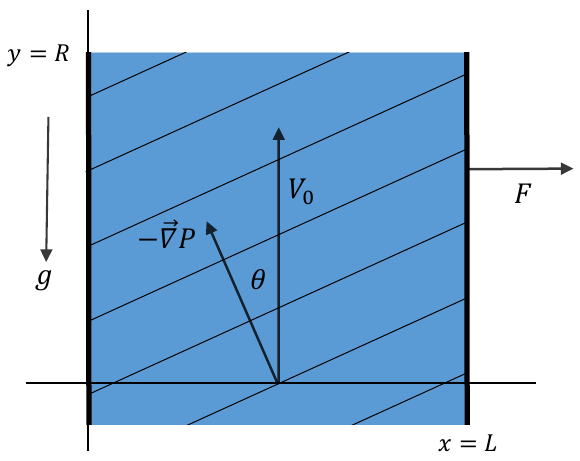}
\caption{Schematic for single fluid channel flow. The fluid flows upward at an angle $\theta$ relative to the pressure gradient, and a transverse force is exerted on the channel. A boundary layer of size $h$ exists on each wall that provides the necessary vorticity to enforce the true no-slip boundary conditions.}
\label{fig:schematic2}
\end{figure}

A simple example that highlights the parity odd behavior is that of a fluid flowing in an infinite channel, $x\in[0,L]$, in the presence of gravity (see Fig~\ref{fig:schematic2}). In this case the analytic velocity $V$ must be constant \footnote{Any bounded analytic function in the infinite strip has to be a constant by an extension of Liouville's theorem.}, since the constant function is the only bounded analytic function over the whole domain. Imposing the no-penetration condition at the walls, that is,
\be
V_x\Big|_{x=0}=V_x\Big|_{x=L}=0,
\ee
we find that $V=-iV_0$, where $V_0$ is a real constant. From Eq.~(\ref{eq:Darcy-complex}), we see that 
\be
W=i\rho g \zeta+i\mu V_0\zeta +W_0,
\ee
where $W_0$ is a complex constant. 

In order to calculate the net force on the sample, we must remember that the original problem is three-dimensional. However, given that $T_{ij}\approx -P\delta_{ij}$ to leading order, and that $P$ does not depend on $z$, the \emph{net force on the sample walls is:}
\be
F_i=\delta_{xi}\, h\int_{-\frac{R}{2}}^{\frac{R}{2}}  P\Big|_{x=0}^L dy=2\beta \, V_0(RLh)\delta_{xi}, \la{eq:force-walls}
\ee
where $R$ is assumed to be the total length of the sample. This approximation is only valid when $R\gg L$. Thus, driving a fluid with parity-odd viscosity through a channel can impart a net force, proportional to the sample volume, in the direction perpendicular to the flow. This is a measurable effect realizable in a laboratory. Even though the origin of this net force may sound mysterious, there is a nice interpretation in terms of 2D electronic fluids. In the presence of magnetic field, a constant flow in the $y$ direction is only possible if the electric field posses a component on the $x$ direction. This is the source of such a net force on the walls.

Since the true boundary condition is the no-slip condition, we can estimate the boundary layer corrections to the net force~(\ref{eq:force-walls}), simply by considering that
\be
\eta_{ixkl} \,\p_k v_l\Big|_{x=0}^L\approx \eta_{ixxy} \,\p_x v_y\Big|_{x=0}^L\sim \eta_{ixxy}\frac{V_0}{h}\,,
\ee
where we assumed that the boundary layer thickness is of order $h$. Comparing this to the pressure, which scales as $h^{-2}$, we see that the boundary layer contribution comes as a higher order correction.

\subsection{Force on an obstacle with arbitrary cross section}

In this section, we consider flow past a cylindrical obstacle with arbitrarily shaped cross section $\mathcal D$ within the HS setup. We show that there are no corrections to the force acting on the compact obstacle coming from the parity breaking terms. The total force on this compact solid body is given by 
\be
F_i=-h\int_0^\ell P \,n_i \,ds, 
\ee
where $ds$ is the arc length element, $\ell$ is the perimeter of the cross section $\mathcal D$ and $\hat n$ is the normal vector pointing outwards from the body. Assuming a positive orientation, the complex normal vector $n_x+in_y$ is given by $-id\zeta/ds$, since $|d\zeta/ds|=1$ by the definition of arc length. Therefore,
\begin{align}
F&\equiv F_x+iF_y=-h\int_0^\ell P\left(-i\frac{d\zeta}{ds}\right)ds\,, \nonumber
\\
F&=ih\varointctrclockwise_{\p\mathcal D} P \,d\zeta\,. \la{eq:complex-force}
\end{align}

For potential flows, we can always define the analytic velocity to be of the form
\be
V=\frac{d\Phi}{d\zeta}\,,
\ee
where the complex potential $\Phi$ is the defined in terms of the velocity potential $\varphi$ and stream function $\psi$ by
\be
\Phi(\zeta)=\varphi(x,y)+i\psi(x,y)\,.
\ee
From~(\ref{eq:Darcy-complex}), we have that 
\be
P=\mbox{Re}\left(-\mu\Phi+i\rho g\zeta\right)+P_0\,,
\ee
where $P_0$ is a real constant. Plugging this into Eq.~(\ref{eq:complex-force}), we end up with
\be
F=\frac{ih}{2}\varointctrclockwise_{\p\mathcal D}\left(-\mu\Phi-\bar\mu\bar\Phi-i\rho g\bar\zeta+i\rho g \zeta+2P_0\right)d\zeta\,. \la{eq:force2}
\ee
We can see that the last two terms vanish by Cauchy's integral theorem. Let the equation for the curve $\p\mathcal D$ be of the form $\bar\zeta=f(\zeta)$. Then
\be
F=-\frac{ih}{2}\varointctrclockwise_{\p\mathcal D}\left(\mu\Phi(\zeta)+\bar\mu\bar\Phi (f(\zeta))\right)d\zeta+\frac{\rho gh}{2}\varointctrclockwise_{\p\mathcal D} f(\zeta) d\zeta\,.
\ee
Before proceeding, let us note that the last term is nothing but the buoyant force on the body. To see this, we use that 
\be
\frac{\rho gh}{2}\varointctrclockwise_{\p\mathcal D}f(\zeta)d\zeta=\frac{\rho gh}{2}\iint_{\mathcal D}d\bar\zeta\, d\zeta=i\rho g\, h\,\mathcal A\,,
\ee
where $\mathcal A\ $ is the area of region $\mathcal D$. This means that this force is always opposite to the gravitational force, and is proportional the mass of fluid displaced by the body, $\rho\times( h \mathcal A )$. 

In order to determine $\Phi$ and $\bar\Phi$ on the curve $\p\mathcal D$, we must impose that $V_n=0$ on $\p\mathcal D$. This implies that
\begin{align}
V_n&=\mbox{Re}\left(-i\frac{d\zeta}{ds}V\right)_{\p\mathcal D},\nonumber
\\
V_n&=-\frac{i}{2}\frac{d\zeta}{ds}\frac{d}{d\zeta}\left[\Phi-\bar\Phi\Big|_{\bar\zeta=f(\zeta)}\right]=0\,.
\end{align}
This shows that the curve $\p\mathcal D$ is a streamline. In other words, $\psi(x,y)|_{\p\mathcal D}$ is a constant. Using this, the \emph{force on a cylindrical body is given by:}
\be
F=-i h \alpha \varointctrclockwise_{\p\mathcal D}\Phi(\zeta) d\zeta+i\rho g h \mathcal A\,. \la{eq:force-thm}
\ee

Note that the first term in Eq.~(\ref{eq:force-thm}) is the drag force, and does not depend on the parity-odd coefficient $\beta$. The reason for that is somewhat straightforward. If we define $\tilde P= P+\beta \psi$, with $\psi$ being the stream function, Eq.~(\ref{Darcy-anis}) can be written as (assuming cylindrical symmetry)
\be
\p_a(\tilde P+\rho g y)=-\alpha v_a\,,
\ee
which has the same form as the ordinary Darcy's law. Moreover, since $\psi(x,y)|_{\p\mathcal D}$ is constant, 
\be
\varointctrclockwise_{\p\mathcal D}P d\zeta=\varointctrclockwise_{\p\mathcal D}\tilde P d\zeta\,,
\ee
and so $\beta$ does not contribute to the drag force. In fact, this true even in the anisotropic case, as shown in Appendix~\ref{sec:aniso}. It should be noted, however, that even though the drag force is unchanged, for most contours the resulting flow pattern will be modified due to $\beta$. This is similar to 3D flows with parity odd terms, where a Stokeslet analysis is seen to modify the flow~\cite{vitelli2021long}.

In the case of an infinitely long channel, it is possible to fully determine the drag force on a cylindrical body. For that we use that the complex potential for a flow with constant complex velocity $U$ at infinity is given by
\be
\Phi(\zeta)=U\zeta+\bar U f(\zeta).
\ee
Plugging this into Eq.~(\ref{eq:force-thm}) gives
\be
F=\left(2\alpha \bar U+i\rho g\right)h \mathcal A\,
\ee
and, because $\bar U=U_{x}+iU_{y}$, we obtain that the drag force is proportional to both the asymptotic fluid velocity, and the volume of the body.

\subsection{Compact free surface problem}

In this section, we will consider the famous HS free surface (moving boundary) problem with parity odd fluids.  We first provide a quick (and incomplete) recap of free surface problems considered for standard HS flows with isotropy. The isotropic free surface problem has been studied since the early work of Galin~\cite{galin1945unsteady} and Polubarinova-Kochina~\cite{polubarinova1945problem}, which was followed up by several authors and has been an active area of research in the form Laplacian growth problem. For a nice review see Howison~\cite{howison1992complex} and the references therein. 

The simplest free surface or moving boundary problem is that of one phase, zero surface tension systems, with the pressure $P(x,y,t)$ satisfying $\Delta P=0$ in the region $\Omega(t)$ occupied by the liquid. The boundary conditions at the free surface $\p \Omega(t)$ are $P|_{\p \Omega(t)}=0$ and $V_n=-\alpha\p_n P$. For the single phase case, the viscous fluid forms a boundary with an inviscid fluid such as air, and the free surface equation coincides with the zero pressure boundary condition, and $V_n=-(\p P/\p t)/|\nabla P|$ is the normal velocity of $\p \Omega(t)$ in the outward direction. The resulting kinematic boundary condition for the free surface $P|_{\p \Omega(t)}=0$ can be written as
   \begin{align}
   	\frac{\p P}{\p t}-\frac{1}{\alpha}|\vec\nabla P|^2=0.
   \end{align}
The above equation can be written in an analytic form using $P=\mbox{Re}(W)$, and then conformally mapped to a unit disc,
\begin{align}
\mbox{Re}\left(\frac{\partial W}{\partial t} - \frac{1}{\alpha}\left\vert\frac{d W}{d \zeta}\right\vert^2\right) =0.
\end{align}
This conformally mapped moving free surface equation is sometimes referred to as Polubarinova-Galin, or Laplacian growth, equation~\cite{howison1992complex}. 
 
\begin{figure}
\centering
\includegraphics[scale=0.97]{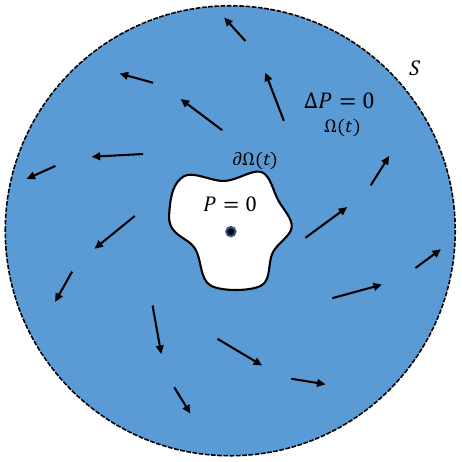}
\caption{Expanding air bubble surrounded by an parity broken viscous fluid in a HS cell. The pressure profile and shape of the interface are unchanged due to $\beta$, however the flow has a spiral behavior.}
\label{fig:schematic5}
\end{figure}
   
     The HS free surface problem is then completely specified when we prescribe a driving mechanism and an initial shape profile $\p \Omega(0)$.   There are two variants of the free surface problem that are often considered: i) the viscous fluid occupies only a finite area surrounded by an inviscid fluid such as air, and the motion is driven by sources or sinks within the viscous fluid, and (ii) a simply connected bubble formed by injection of an inviscid fluid, for example air, into an infinite region of a second fluid whose viscosity is large. These two problems can be framed in a similar way, but there are key differences in the dynamics with respect to the stability of the free surface. 
     
     Here we only consider the case where the parity broken viscous fluid occupies the exterior $\Omega(t)$ of the finite bubble, with uniform extraction at infinity. If the air is injected at a rate $q$ given by
   $q = \frac{d\mathcal A}{dt}$, where $\mathcal A$ is the area, then far from the injection point and bubble the outer fluid has a solution for $W$ of the from
\begin{align}
    \lim_{r\rightarrow \infty}W(\zeta) \sim - \frac{q}{2\pi}\log(\zeta) + W_0 .
\end{align}
The pressure profile cannot have angular dependence, otherwise it would not be a single valued function. From~(\ref{eq:Darcy-complex}), the far field complex velocity must then be of the form
\begin{align}
    V(\zeta) \sim \frac{q}{2\pi\mu}\frac{1}{\zeta} = \frac{q}{2\pi\mu} \frac{e^{-i\theta}}{r} \la{eq:farflow} .
\end{align}
The radial and angular components of the far field velocity are then
\begin{align}
    V_r \sim \frac{q}{2\pi}\frac{\alpha}{\alpha^2+\beta^2}\frac{1}{r},\quad 
    V_{\theta} \sim \frac{q}{2\pi}\frac{\beta}{\alpha^2+\beta^2}\frac{1}{r},
\end{align}
and so it's clear that $\beta$ causes the flow to acquire circulation. The circulation and flux can be computed far from the bubble. Let $S$ be a curve far from the bubble, such that the the velocity is described by~(\ref{eq:farflow}). Then
\begin{align}
    \varointctrclockwise_{S} V(\zeta) d\zeta \sim i\frac{q}{\mu}.
\end{align}
\emph{The flux $\Phi_v$ and circulation $\Gamma$ at infinity are then given by:}
\begin{align}
    \Phi_v  \sim \frac{\alpha}{\alpha^2+\beta^2}\frac{d\mathcal A}{dt},\quad 
    \Gamma  \sim \frac{\beta}{\alpha^2+\beta^2}\frac{d\mathcal A}{dt}.
\end{align}
Thus, if the area of the air bubble is changing, the presence of circulation at the edge of the sample can be used to measure the parity broken terms in the viscosity tensor.  

For a free surface parametrized by $P|_{\p \Omega(t)}=0$, the outward normal unit vector is given by $\hat n=\frac{\vec\nabla P}{|\vec\nabla P|}$, and so the kinematic boundary condition can be written as
\begin{align}
		\frac{\p P}{\p t}+V_n |\vec \nabla P|=0.
		\label{eq:kbc}
\end{align}
For parity odd fluids we have the following $\beta$ modified form for $V_n$,
\begin{align}
	V_n=-\frac{1}{\alpha^2+\beta^2}\left(\alpha \partial_n P+\beta \partial_s  P\right) .
\label{eq:vn}
\end{align}
Since $\hat n=\frac{\vec\nabla P}{|\vec\nabla P|}$, the second term in the above equation vanishes. Substituting Eq.~(\ref{eq:vn}) into Eq.~(\ref{eq:kbc}), we obtain,
\begin{align}
  \left(  \frac{\partial P}{\partial t}-\frac{\alpha}{\alpha^2+\beta^2}\vert\Vec{\nabla}P \vert^2\right) \Big|_{\p \Omega(t)}=0. \la{eq:Lapgrow}
\end{align}
From this it's clear that the shape dynamics of the bubble are effectively unchanged due to $\beta$, apart from rescaling time by $\tilde \alpha$,
\begin{align}
    \frac{1}{\tilde{\alpha}} = \frac{\alpha}{\alpha^2+\beta^2}. \la{eq:rescale}
\end{align}
Substituting $P=\mbox{Re}(W)$ and conformally mapping $\Omega(t)$ to a unit disc yields the P-G equation with renormalized rate $q$, 
\begin{align}
\mbox{Re}\left(\frac{\partial W}{\partial t} - \frac{1}{\mu}\left\vert\frac{d W}{d \zeta}\right\vert^2\right) =0.
\end{align}

Typically, the growth of the bubble is limited by a critical `blow-up' time $t_c$, which is the time it takes for sharps cusps to form~\cite{howison1992complex}. Since the shear viscosity is a determining factor in $t_c$, the rescaling in~(\ref{eq:rescale}) implies that $\beta$ would modify $t_c$. In particular, large values of $\beta$ would delay this cusp formation for a fixed rate $q$.
We would like to point out that the system studied here is closely related to the free surface dynamics in a rotating HS cell except for one important difference. In the rotating case, the centrifugal force modifies the pressure at the boundary, resulting in a different equation for the surface dynamics~\cite{schwartz1989rotate, alvarez2008coriolis}.

\subsection{Dispersion and stability of the free surface interface between two fluids: Saffman-Taylor instability}

In this section we study the Saffman-Taylor instability in the presence of the parity breaking terms and derive a modified free surface dispersion relation and stability condition. Consider a setup with two superposed fluids subject to a downward gravitational force acting in the negative $y$ direction, with an interface between them moving with speed $V_0$. In the following analysis $V_0$ can be positive or negative; positive values correspond to pumping the fluid in the positive $y$ direction, and negative values correspond to pumping the fluid in the negative $y$ direction.

At a particular instant in time the unperturbed interface between the two fluids is located at $y=0$, and perturbations assumed to be of the from
\begin{align}
y= H(x,t)  = \epsilon\,\mbox{Re}\left( e^{i k x+\Omega t}\right), \la{eq:inter}
\end{align}
where $\epsilon$ is the small amplitude of the perturbation. We have also allowed for the possibility of a complex valued frequency $\Omega = \delta +i \omega$. All quantities associated with the upper fluid ($y>0$) are marked with a $1$, and all quantities associated with the lower fluid ($y<0$) are marked with a $2$. The complex viscosity that enters~(\ref{eq:Darcy-complex}) in each region is denoted by  $\mu_1 = \alpha_1+i\beta_1$ and $\mu_2 = \alpha_2+i\beta_2$.

\begin{figure}
\centering
\includegraphics[scale=0.9]{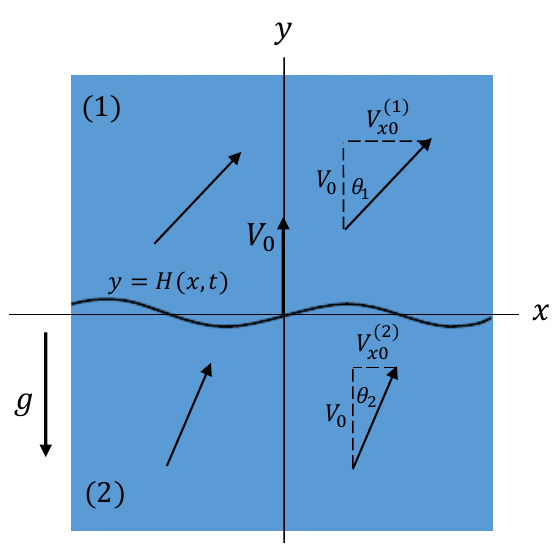}
\caption{Interface between two odd fluids moving with speed $V_0$ upwards, with small amplitude perturbations. The asymptotic velocity in region (1) makes an angle $\theta_1$ with the vertical, and the asymptotic velocity in region (2) makes an angle $\theta_2$ with the vertical. If $\beta=0$ both angles can be set to zero.}
\label{fig:schematic3}
\end{figure}

We start with general solutions for $P$ in each region that satisfy Laplace's equation, and compute the corresponding components of the flow using~(\ref{eq:Darcy-complex}). We then apply the kinematic boundary condition at the free surface,
\begin{align}
V_y^{(1)}-V_x^{(1)} \partial_x H(x,t) = V_0 + \partial_t H(x,t).
\end{align}
and impose the boundedness of the flow at $y \rightarrow \pm \infty$. To first order in $\epsilon$, we have
\begin{align}
P^{(1)}(x,y) = P_0^{(1)} + a_1 x + b_1 y \quad\quad\quad\quad\quad\quad\quad\quad \nonumber \\
- \epsilon e^{-k y +\delta t}\left(A_1\cos(kx+\omega t)+B_1\sin(kx+\omega t)\right),\la{eq:pressure1} \\
P^{(2)}(x,y) = P_0^{(2)} + a_2 x + b_2 y \quad\quad\quad\quad\quad\quad\quad\quad \nonumber \\
- \epsilon e^{k y +\delta t}\left(A_2\cos(kx+\omega t)+B_2\sin(kx+\omega t)\right), \la{eq:pressure2}
\end{align}
where  $P_0^{(1)}$ and $P_0^{(2)}$ are constant background pressures in each region, and where
\begin{align}
a_i &= -\beta_i V_0 - \alpha_i V_{x0}^{(i)},\\
b_i &= -\alpha_i V_0 + \beta_i V_{x0}^{(i)},
\end{align}
are the constants determining the steady state background flow, and
\begin{align}
A_1 &= \frac{\beta_1(\omega+k V_{x0}^{(1)})-\alpha_1\delta}{k},\\
B_1 &= \frac{\alpha_1(\omega+k V_{x0}^{(1)})+\beta_1\delta}{k},\\
A_2 &= \frac{\beta_2(\omega+k V_{x0}^{(2)})+\alpha_2\delta}{k},\\
B_2 &= \frac{-\alpha_2(\omega+k V_{x0}^{(2)})+\beta_2\delta}{k},
\end{align}
are the amplitudes of perturbation. The constants $V_{x0}^{(1)}$ and $V_{x0}^{(2)}$ are the $x$ components of fluids as $y \rightarrow \infty$ and $y \rightarrow -\infty$, respectively (see Fig~\ref{fig:schematic3}). At this point we have not fixed the asymptotic form of the velocities, we only require that they agree kinematically with the interface to order $\epsilon$.

We must also balance the forces at the interface. In general, the no-stress boundary condition requires that $n_i T_{ij}$ be continuous across the free surface, and this continuity must be verified across any boundary layer that develops. In the standard HS cell there exists a boundary layer flow that interpolates between the tangential velocity of each fluid on either side of a free surface. Along with the HS scaling, this leaves only a single effective boundary condition on bulk solutions, that the pressure must be continuous from one region to the next:
\begin{align}
P^{(1)}(x,y)=P^{(2)}(x,y), \quad y = H(x,t) .
\end{align}
In the standard case, where $\beta=0$, this leads to a jump in the bulk scale tangential velocity near the free surface, but does not place any constraints on the asymptotic flow. However, the introduction of $\beta$ modifies this jump condition, and does in fact place a constraint on the flow. This manifests itself in the form of the asymptotic velocity~(\ref{eq:S2}).

Upon substituting~(\ref{eq:inter}) into (\ref{eq:pressure1}) and (\ref{eq:pressure2}) and setting them equal, we are given four equations, two at order $\epsilon^0$,
\begin{align}
    P_0^{(1)} & =P_0^{(2)} , \la{eq:S1}\\
    \alpha_1 V_{x0}^{(1)} + \beta_1 V_0 &= \alpha_2 V_{x0}^{(2)} + \beta_2 V_0 , \la{eq:S2}
\end{align}
and two at order $\epsilon^1$,
\begin{align}
      (\alpha_1+\alpha_2)\delta-(\rho_1-\rho_2)g k \quad\quad\quad\quad\quad\quad\quad\quad\quad\quad\quad\quad \nonumber \\
      -(\alpha_1-\alpha_2)V_0k-(\beta_1-\beta_2)\omega=0, \quad\la{eq:S3} \\
      (\alpha_1+\alpha_2)\omega+(\alpha_1 V_{x0}^{(1)}+\alpha_2 V_{x0}^{(2)})k+(\beta_1-\beta_2)\delta=0. \la{eq:S4}
\end{align}
Eq.~(\ref{eq:S1}) implies the constant background pressures must be equal on both sides, while Eq.~(\ref{eq:S2}) gives a constraint on the asymptotic form of the flow. If $\beta_1-\beta_2=0$ the fluid can be pumped purely in the vertical direction. However, with $\beta_1-\beta_2 \neq 0$, the fluid being driven (fluid 1, say) moves at an angle relative to the driving fluid (fluid 2, say). If we introduce an asymptotic flow angle in each region defined by
\begin{align}
   V_{x0}^{(1)} &= V_0 \tan\theta_1,\\
   V_{x0}^{(2)} &= V_0 \tan\theta_2,
\end{align}
then~(\ref{eq:S2}) can be written as
\begin{align}
    \alpha_1 \tan\theta_1 = \alpha_2 \tan\theta_2 + (\beta_2-\beta_1).
\end{align}
This implies that for $\beta_1-\beta_2 \neq 0$, two superposed fluids cannot be pumped in the purely vertical direction. Said another way, there must be an angle between the steady state flow and the free surface.

The system of equations~(\ref{eq:S3}) and~(\ref{eq:S4}) can be used to solve for $\delta$ and $\omega$, and when written as a single complex number we arrive at the\emph{ modified free surface dispersion relation:}
\begin{align}
\frac{\Omega}{k} = \frac{(\rho_1-\rho_2)g + V_0(\mu_1-\mu_2)-2i\alpha_2 V_0 \tan\theta_2}{\mu_1+\bar{\mu_2}}. \la{eq:modisp}
\end{align}
If $\beta \rightarrow 0$, and the flow is normally incident to the interface in region (2), that is $\theta_2=0$, then the frequency $\Omega$ becomes a real number, and the standard Saffman-Taylor dispersion is recovered~\cite{saffman1958taylor}. For concreteness, we set $\theta_2=0$ from here onward, as to model a fluid being driven directly at the interface. In this case, Eq.~(\ref{eq:modisp}) is simply the complexified version of the Saffman-Taylor dispersion, where the viscosities have been replaced by their complex generalizations.

While the modified dispersion is still linear in $k$, the conditions for stability are changed due to $\beta$. Stability occurs when $\mbox{Re}\left(\Omega\right)=\delta<0$, giving the \emph{modified free surface stability condition:}
\begin{align}
(\rho_1-\rho_2)g+V_0(\alpha_1-\alpha_2)+\frac{V_0(\beta_1-\beta_2)^2}{\alpha_1+\alpha_2} <0 . \la{eq:stability}
\end{align}
Written another way,
\begin{align}
    V_0\left(\alpha_1-\alpha_2 + \frac{(\beta_1-\beta_2)^2}{\alpha_1+\alpha_2}\right) < (\rho_2-\rho_1)g ,
\end{align}
we can see that for fixed densities, stability depends on the sign of the quantity
\begin{align}
    \alpha_1-\alpha_2 + \frac{(\beta_1-\beta_2)^2}{\alpha_1+\alpha_2},
\end{align}
which ultimately depends on the relative values of $\alpha_1$ and $\alpha_2$ in relation to $\beta$. In the absence of any parity broken terms, we recover the Saffman-Taylor stability condition.

The typical case in which the interface is unstable is that of a dense, viscous fluid resting on top of a less dense, less viscous fluid ($\rho_1>\rho_2$, $\alpha_1>\alpha_2$). In this case, stability is achieved when
\begin{align}
    V_0 < -\frac{(\rho_1-\rho_2)(\alpha_1+\alpha_2)g}{\alpha_1^2-\alpha_2^2 +(\beta_1-\beta_2)^2}. \label{eq:vcond}
\end{align}
From this we see that the parity broken terms in the viscosity tensor act to stabilize the interface. Typically the fluid is required to be pumped downward with sufficiently negative $V_0$, however the extra factor of $\left(\beta_1-\beta_2\right)^2$ in the denominator implies the velocity does not need to be as negative. In the extreme situation, when $\alpha_1=\alpha_2=\alpha$, the modification is most striking. In this case $\beta$ is the only way to reintroduce $V_0$ into the stability condition, and stability condition (\ref{eq:vcond}) becomes
\begin{align}
    V_0 < \frac{2(\rho_2-\rho_1)g\alpha}{(\beta_1-\beta_2)^2}.
\end{align}
This opens an entirely new channel for stability that wasn't present when $\beta=0$.

Regardless if the fluid is driven or not, the introduction of $\beta$ gives the perturbed interface a time dependent oscillation, with a frequency $\omega$ proportional to $\big\vert\beta_1 - \beta_2 \big\vert$. Stable configurations return to equilibrium as a damped oscillator, and the unstable configurations oscillate with ever increasing amplitude. This is a directly measurable quantity, and acts as a robust probe into the parity broken terms in the viscosity tensor.

\section{Discussion and Future directions} \la{sec:disc}

In this work, we derived the flow equations for a three-dimensional incompressible fluid with a general parity broken anisotropic viscosity tensor, when placed between two parallel plates with a small seperation $h$. In the infinitesimal gap limit, Darcy's law admits a simple generalization that contains only four viscosity coefficients, as shown in Eq.~(\ref{eq:gen-Darcy}). We discussed the observable effects of the parity odd coefficients (for a cylindrical symmetric case) of the fluid in a channel flow, flow around an obstacle, expanding bubble, and two-fluid interface stability (Saffman-Taylor instability).

When such a fluid is pushed through a channel, a transverse force is exerted on the walls due to the parity odd coefficients. Measurement of this transverse force can enable us to determine the magnitude of such parity odd coefficients in both synthetic and naturally occurring three dimensional fluids. For a flow across an obstacle, the drag force is independent of the parity breaking effects, which is in contrast to recent results in three-dimensional systems with odd viscosity~\cite{vitelli2021long}. In the case of an expanding bubble, the pressure profile and interface dynamics are unchanged, however there is a modification to the far field flow, and measurement of the fluid circulation far from the bubble gives a measure of the parity odd terms. The stability condition of the two fluid interface is also modified due to the presence of parity breaking, with the parity breaking tending to stabilize the interface dynamics. 

In principle, the parity odd behavior presented here could arise from a magnetized fluid (colloid) subject to a uniform external magnetic field along the $z$ axis. This assumes the fluid is incompressible and satisfies the Laundau-Lifshitz equation~\cite{landau1935theory},
\be
D_t\boldsymbol M=-\chi \boldsymbol M\times\boldsymbol B-\lambda  \boldsymbol M\times(\boldsymbol M\times\boldsymbol B)\,,
\ee
together with
\be
\rho D_t v_i=\p_jT_{ij},
\ee
where $D_t=\p_t+v_j\p_j$ is the material derivative, and the stress tensor is given by
\begin{align}
T_{ij}=&-P\delta_{ij}+\eta\,(\p_iv_j+\p_jv_i) \nonumber
\\
&+ \nu M_k\left[\epsilon_{jkl}(\p_iv_l+\p_lv_i)+\epsilon_{ikl}(\p_jv_l+\p_lv_j)\right].
\end{align}
The above set of equations are the minimal model that yields the desired parity breaking effects resulting from the relaxation dynamics of the magnetization equation. The above equations resemble the three dimensional fluids discussed in Ref.~\cite{markovich2020odd}, where the magnetization plays the role intrinsic angular momentum, albeit with Landau-Lifshitz dynamics. 

An interesting question for the future is to investigate if these equations can arise in Ferrofluids, or their close counterparts, dipolar fluids (see Ref.~\cite{rosensweig2013ferrohydrodynamics, rosensweig1983labyrinthine}). Ferrofluids seem to be a promising platform to study the parity breaking fluids discussed in this work, since they manifest remarkable features, such as labyrinthine instabilities, when placed within a Hele-Shaw device~\cite{rosensweig1983labyrinthine, jackson1994hydrodynamics}.

\section{Acknowledgments} 

We thank A.G. Abanov for useful discussions. This work is supported
by NSF CAREER Grant No. DMR-1944967 (SG).  DR is supported by 21st century foundation startup award from CCNY. GMM was supported by the National Science Foundation under Grant OMA-1936351

\begin{appendix}

\section{Fully anisotropic case} \la{sec:aniso}

Even though the bulk of our analysis was done using cylindrical symmetry, the anisotropic case with arbitrary matrix elements $\mathfrak{y}_{ij}$ can be shown to acquire a similar complex generalization, albeit with modified analytic functions and boundary conditions. To derive a complex form of the Hele-Shaw flow equations for the anisotropic case, 
it is convenient to introduce the isothermal coordinates
\be
\sigma=\sqrt{\gamma_{xx}}\,x+\frac{\gamma_{xy}}{\sqrt{\gamma_{xx}}}\,y\,, \qquad \uptau=\frac{y}{\sqrt{\gamma_{xx}}}\,, \la{isothermal}
\ee
such that, in this new coordinate system, Eq.~(\ref{Darcy-anis}) and (\ref{eq:incompress2}) become
\begin{align}
    &\p_\sigma\Big(P+\rho g\uptau\sqrt{\gamma_{xx}}\Big)=-\alpha\, V_\sigma-\beta\, V_\uptau\,, \la{Darcy-sigma}
    \\
    &\p_\uptau\Big(P+\rho g\uptau\sqrt{\gamma_{xx}}\Big)=-\alpha \,V_\uptau+\beta\, V_\sigma\,, \la{Darcy-tau}
    \\
    &\p_\sigma V_\sigma+\p_\uptau V_\uptau=0\,, \la{incomp-isothermal}
\end{align}
where we have defined
\be
V_\sigma=\sqrt{\gamma_{xx}}\,V_x+\frac{\gamma_{xy}}{\sqrt{\gamma_{xx}}}\,V_y\,, \qquad V_\uptau=\frac{V_y}{\sqrt{\gamma_{xx}}}\,.
\ee
Equations~(\ref{Darcy-sigma}-\ref{incomp-isothermal}) imply that  
\be
(\p_\sigma^2+\p_\uptau^2)P=0\,, \qquad \p_\sigma V_\uptau-\p_\uptau V_\sigma=0\,,
\ee
that is, pressure is a harmonic function in these new coordinates and the function $V=V_\sigma-iV_\uptau$ is analytic, since it satisfies the Cauchy-Riemann equations in the isothermal coordinates. Moreover, since $P(\sigma,\uptau)$ is a harmonic function, we can always define a function $Q(\sigma,\uptau)$ such that $W=P+i\,Q$ is analytic. In terms of the complex variables $\zeta= \sigma+i\,\uptau$, $V$ and $W$, equations~(\ref{Darcy-sigma}-\ref{incomp-isothermal}) can be written as
\be
\frac{d}{d\zeta}\Big(W-i\rho g\sqrt{\gamma_{xx}}\,\zeta\Big)=-\mu V(\zeta), \la{eq:Darcy-complex-anis}
\ee
where $\alpha$ and $\beta$ are combined into the complex viscosity $\mu=\alpha+i\beta$. 

Analogous to the cylindrical symmetry case, only $\alpha$ contributes to the drag force. To see this, we must express the drag force on the body in terms of a contour integral in the complex $\zeta$-plane. Eq.~(\ref{eq:complex-force}) gives us
\be
F_x=-h\,\varointctrclockwise_{\p\mathcal D} P \,dy\,,\qquad F_y=h\,\varointctrclockwise_{\p\mathcal D} P \,dx\,,
\ee
and with the help of Eq.~(\ref{isothermal}), we can write
\be
\tilde F=\sqrt{\gamma_{xx}}\,F_x+\frac{\gamma_{xy}+i}{\sqrt{\gamma_{xx}}}\,F_y=-i h\,\varointctrclockwise_{\p\tilde{\mathcal  D}} P \,d\zeta\,,
\ee
where $\tilde{\mathcal D}$ is the object domain in the complex $\zeta$-plane. Since we are only interested in the drag force, let us ignore the gravity term. From Eq.~(\ref{eq:Darcy-complex-anis}), we obtain that
\begin{align}
P&=\mbox{Re}(W)=\mbox{Re}(\mu\Phi+W_0)\,, \nonumber
\\
P&=\alpha\,\mbox{Re}(\Phi)-\beta\,\mbox{Im}(\Phi)+\mbox{Re}(W_0)\,, 
\end{align}
where $W_0$ is a complex constant. However, one can see that 
\begin{align}
V^i n_i&=\frac{dy}{ds}V_x-\frac{dx}{ds}V_y=\frac{d\uptau}{ds}V_\sigma-\frac{d\sigma}{ds}V_\uptau\,, \nonumber
\\
V^i n_i&=\mbox{Re}\left(-i\frac{d\zeta}{ds}V\right)=\mbox{Im}\left(\frac{d\Phi}{ds}\right)=0\,,
\end{align}
which implies that $\mbox{Im}(\Phi)$ is constant on the contour $\p\tilde{\mathcal D}$. Therefore, we are only left with
\be
\tilde F=-i\alpha h\,\varointctrclockwise_{\p\tilde{\mathcal  D}} \Phi(\zeta) \,d\zeta\,.
\ee
From this, we directly observe that $\beta$ does not contribute to the drag force, even in the anisotropic case, as expected.

\end{appendix}

\bibliographystyle{refstyle}
\bibliography{OddDarcyLaw-Bibliography.bib}

\end{document}